# Self-organization of the Earth's climate system versus Milankovitch-Berger astronomical cycles

Lev A. Maslov

University of Northern Colorado, Department of Physics and Astronomy,

Greeley, CO, 80639, USA

lev.maslov@aims.edu; lev.maslov@unco.edu

**Abstract**

The Late Pleistocene Antarctic temperature variation curve is decomposed into two components: "cyclic" and "high frequency, stochastic". For each of these components, a mathematical model is developed which shows that the cyclic and stochastic temperature variations are distinct, but interconnected, processes with their own self-organization. To model the cyclic component, a system of ordinary differential equations is written which represent an auto-oscillating, self-organized process with constant period. It is also shown that these equations can be used to model more realistic variations in temperature with changing cycle length. For the stochastic component, the multifractal spectrum is calculated and compared to the multifractal spectrum of a critical sine-circle map. A physical interpretation of relevant mathematical models and discussion of future climate development within the context of this work is given.

# 1. Introduction

There are a number of different approaches to modeling Earth's climate dynamics, and the author will discuss three primary methods. In the first approach, the climate is governed by external, astronomical forces. These forces include variations in solar activity, changes in the tilt of the Earth's axis, variability in the distance from the Sun, and other parameters associated with Earth's orbit. This model was popularized by M. Milanković and was first published in Serbian in 1912; see also (Milanković, 1998). Today, the field of climate science is represented by the publications of Andre Berger and his followers (Berger, 1988; Berger and Loutre, 1991; Berger et al, 1992; Berger and Loutre, 2002). The study of the influence of astronomical cycles on the Earth's climate is a Ptolemaic work that consists of combining astronomical cycles, and estimating how the resulting solar radiation affects the Earth. In this approach, the Earth's climate system is considered a black box with variations in solar radiation as the input, but unclear output. No self-sustained dynamical model of climate was developed in this approach. Instead, the search for correlations between input and output parameters is the main method used to study the influence of astronomical cycles on climate dynamics. These correlations are not proof that solar cycles are the dominating factor in climate change. For example, the main peak of the power spectrum of $\delta^{18}O$ variation in Antarctica is represented by a period of approximately 100,000 years. The authors (Muller and McDonald, 1997) conclude that this period is caused by changes in the orbital inclination of the Earth's orbit.

In this work's alternative approach, the climate is represented as a complex multi-component dynamical system. The driving forces in this model are internal rather than external.

The first self-sustained dynamical multicomponent glacier-ocean-atmosphere model of climate was developed by Sergin (Sergin, 1979). A system of differential equations representing this model were developed and solved numerically. It was found that characteristic parameters of the climate system auto-oscillate with periods of 20,000-80,000 years. Another complex self-sustained model of the interaction between continental ice, ocean and atmosphere was developed in 1993 (Kagan et al, 1993). A system of differential equations representing this model was written. Given realistic thermodynamic parameters, the authors found a similar auto-oscillation within this system, with a period of about 100,000 years.

The enormous complexity of climate processes requires the use of methods and models capable of handling such complexity. Implementing the modern theory of dynamical systems brought a breakthrough in understanding and modeling climate dynamics. Thus, the application of multifractal statistics to the study of temperature variations in Greenland (Schmitt et al, 1995) and Antarctica (Ashkenazy et al., 2003) allowed the authors to formulate the requirements for realistic climate models, which must "include both periodic and stochastic elements of climate change". The importance of viewing Earth's climate as a nonlinear, complex, dynamical system was understood in 2004 (Rial, Pielke, Beniston, et al., 2004), but no references were made to the earlier publications (Sergin, 1979; Kagan et al, 1993; Schmitt et al, 1995; Ashkenazy et al., 2003) which modeled climate as a nonlinear, complex, dynamical system, and no mathematical, or conceptual models were suggested.

The approach in the current work is based solely on real data. We study the Antarctic Late Pleistocene temperature record calculated from the hydrogen isotope ratio of Vostok ice-core data, (Petit et al, 1999). Figure 1 shows a graph of the temperature variation in Antarctica over the last 435,000 years.

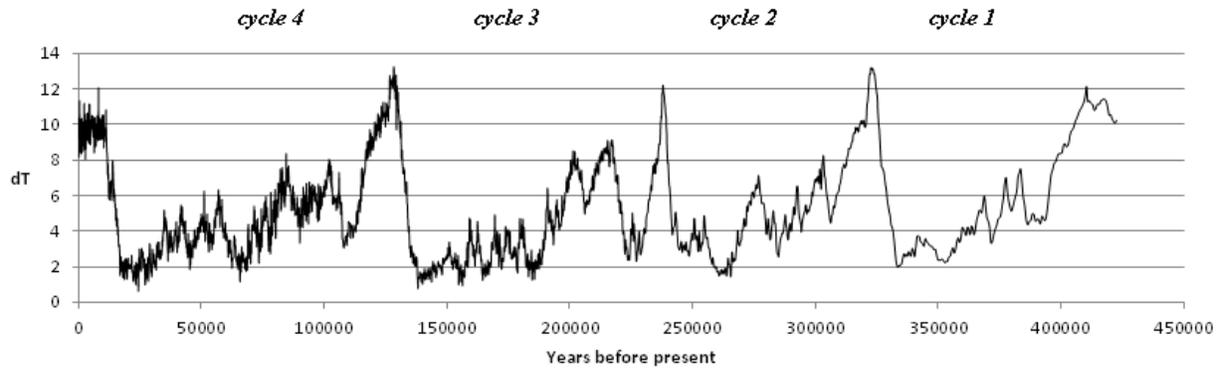

Figure 1

Temperature variations (*ΔT*) in Antarctica according to Vostok ice-core data.

Our goal is three-fold: a) decomposition of the temperature variations (Figure 1) into components corresponding to basic processes, b) mathematical modeling of these processes, c) physical interpretation of these mathematical models and discussion of a consolidated model of climate cycles.

## 2. Decomposing the Data

For the last 420 thousand years the planet has experienced four nearly identical episodes of temperature change. Each episode starts with a sharp increase in temperature, followed by long gradual cooling. On a conceptual level, this can be interpreted as a combination of two subsequent processes: *a)* the release of latent heat stored in the system and *b)* dissipation of this heat by thermal convection. To test the first hypothesis, and to estimate the rise in temperature due to atmospheric water vapor condensation, suppose the enthalpy *H* of the atmosphere remains constant throughout the cycle:

$$H = c_p \cdot T + q \cdot L = const \qquad (1)$$

In this formula $c_p$ is the specific heat of dry air, $T$- temperature, $L$ is the heat of condensation of water vapor, and $q$- is the ratio of the total mass of water vapor in the atmosphere to the total mass of dry air. According to (Trenberth and Smith, 2005) for the current epoch $q = 0.00247$. Variations in temperature $\Delta T$ and of the ratio of the total mass of water vapor in the atmosphere to the mass of dry air $\Delta q$ must compensate each other, such that

$$c_p \cdot \Delta T + \Delta q \cdot L = 0 \qquad (2)$$

For $c_p = 1.006$ J/(g °C), $L = 2500$ J/g, $\Delta q = 0.00247$ (all the water vapor is condensed into water), we find $\Delta T \approx 6.14$ °C. This estimate is reasonably close to the changes in temperature at the beginning of each interglacial cycle.

To test the second hypothesis, we use Newton's Law of Cooling, $dT/dt = sT, s < 0$ which gives us an exponential decrease in the temperature due to thermal convection. The natural logarithm of $T(t)$ is plotted in Figure 2.

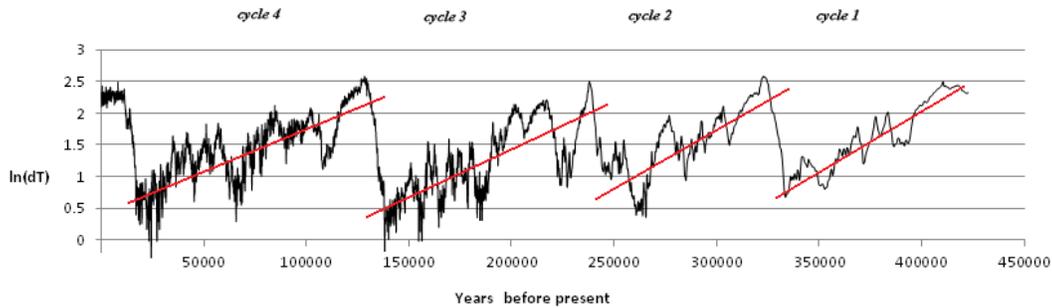

Figure 2

$ln(T)$ function, and straight line approximations $ln(T)=k_i t + p_i$ within cycles.

Next, the data within each cycle is approximated by $ln(T)=k_i t + p_i$ and four exponential functions $T_e(t)=exp(k_i t+p_i)$, $i=1,2,3,4$, are constructed and plotted against the original temperatures, Figure 3a. Figure 3b shows the difference between the observed temperature $T$ and $T_e$.

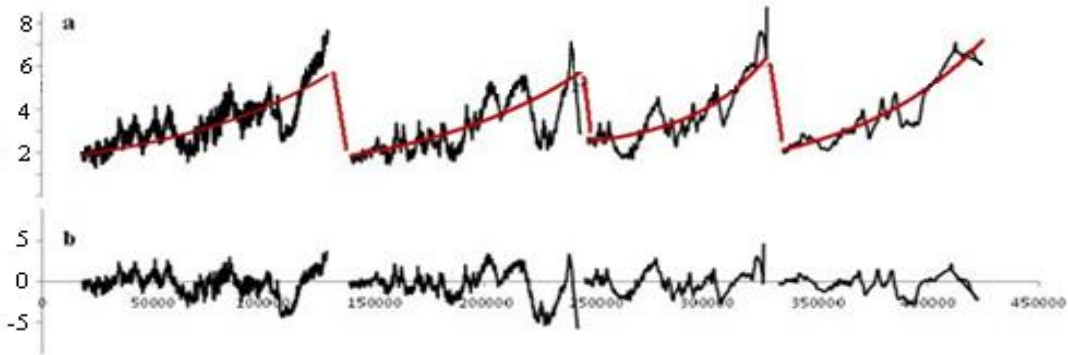

Figure 3

Two components of the temperature variation in Antarctica; *a* - the original data superimposed with exponential functions $T_e(t)=exp(k_i t+p_i)$, $i=1,2,3,4$, for each cycle; *b* – high frequency oscillation components as a difference between the original data and $T_e(t)$.

The coefficient of determination for this approximation was calculated to be $R^2 = 0.624$, which is equivalent to a coefficient of correlation $r = 0.79$.
Thus, the temperature variation in Antarctica can be represented as a sum of two components: the cyclic, low frequency nonlinear oscillation component, and a high frequency stochastic oscillation component. For simplicity, we will refer to these two components as "cyclic" $T_c$ and "high frequency stochastic" $T_s$; $T = T_c + T_s$.
The current approach to decomposing the observed data is based on a physical principle and is different from methods based on formal frequency filtering of the data.

## 3. Modeling the Cyclic Component of Temperature

To model the cyclic component a number of ODE systems were considered - including Lorenz, Brusselator, and Lotka-Volterra equations. After a number of numerical experiments with the systems mentioned above, the Lotka-Volterra equations were modified with variable coefficients and accepted for the current research:

$$\frac{d}{dt}T_c(t) = a_1(t) \cdot T_c(t) + b_1(t) \cdot B(t) \cdot T_c(t)$$
$$\frac{d}{dt}E_c(t) = a_2(t) \cdot E_c(t) + b_2(t) \cdot B(t) \cdot E_c(t) \qquad (3)$$
$$\frac{d}{dt}B(t) = c(t) \cdot B(t) \cdot T_c(t) + d(t) \cdot B(t) \cdot E_c(t)$$

In these equations $T_c(t)$ represents the temperature of a system, $E_c(t)$ – thermodynamic entropy, $B(t)$ – "buffer" function. This function collects internal energy and distributes it between the "thermal" and "entropy" elements of a system.

$$B = C \cdot exp \int_0^{\Delta t} (c \cdot T_c(t) + d \cdot E_c(t))dt \qquad (4)$$

Redistribution of $T_c(t)$ and $E_c(t)$ in the system is governed by $b_1 \cdot B(t) \cdot T_c(t)$ and $b_2 \cdot B(t) \cdot E_c(t)$ terms in the RHS of the first and second equations in (3). In this section we consider equations (3) with constant coefficients, and in section 5 we will give an example of a solution with variable coefficient $a_1(t)$. Applying the methods of nonlinear dynamics (Hilborn, 2006) we found that system (3) has two stationary points. The eigenvalues of the Jacobian matrix for the first point are: $\lambda_1 = a_1, \lambda_2 = a_2, \lambda_3 = 0$. For the second stationary point the eigenvalues are:

$\lambda_1 = 0; \lambda_2, \lambda_3 = \pm\sqrt{BdE(b_1 - b_2)}$. For some combination of coefficients, say for $b_1 > b_2, d < 0$, the eigenvalues $\lambda_{2,3}$ are purely imaginary which results in the origin of a cycle. Figure 4 gives an example of modeling the cyclic component of temperature variations in Antarctica by a solution of a system of equations (3) for coefficients $a_1$= -4.673, $b_1$=0.1827, $a_2$=17.9005, $b_2$= -0.70, c= -2.0, d=3.5, and for initial conditions **$T_c(0)$=8, $E_c(0)$=0.00015, $B(0)$=150.**
Numerical integration of (3) was made in Maple using the Runge-Kutta method on a uniform time scale with the number of steps proportional to 430ky. The theoretical curve, Figure 4b, represents the periodic oscillation of temperature with period ≈100ky.

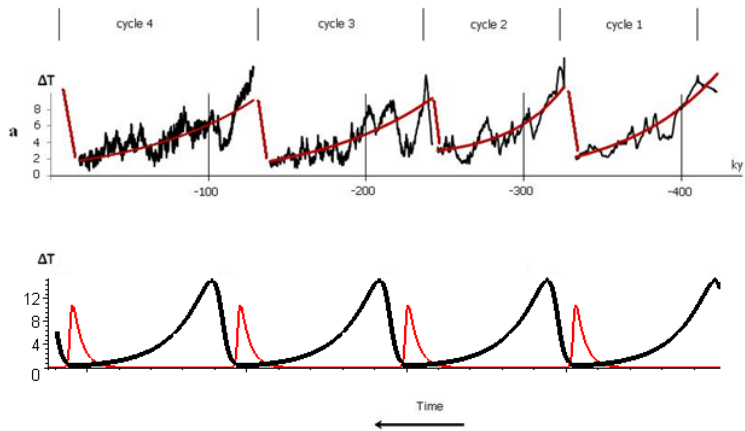

Figure 4

Modeling temperature component **$T_c$**.

**a** – The original data; **b** – Calculated temperature variation

**$T_c(t)$** from equations (3), and corresponding variation of entropy (red).

One unit of computational cycle on the time scale corresponds to 100ky.

## 4. Multiscale Statistics of the High Frequency Part of the Temperature Curve

Study of the multifractal structure of the Greenland ice-core climate proxy temperatures $\delta^{18}O$ is presented in (Schmitt et al, 1995). It is shown here that the power spectrum of the data has the form $S(f) \approx f^{-\beta}$ with *β=1.4.* It is also shown that the data set is a multifractal - i.e. a collection of monofractal subsets with individual scaling exponents. Multifractals are effective in describing and explaining many complex natural phenomena such as thermal convection in fluid dynamics, solar activity, earthquake phenomenology, and more (Harte, 2001). In the work (Ashkenazy et al, 2003) the isotopic temperature record from the Vostok, Antarctica, ice-core is studied. The authors calculated the mass exponent *τ(q)* from time series *T(t)* and showed that this spectrum is nonlinear for $-5 < q < 5$ interval of moments. The authors formulated the requirements for realistic climate models which must "include both periodic and stochastic elements of climate change". Neither detrending, nor decomposition of the original data, similar to that done in section 2 of the current work, were reported in the publications mentioned above. The high frequency component of the temperature in cycles 1-4 is shown in Figure 5.

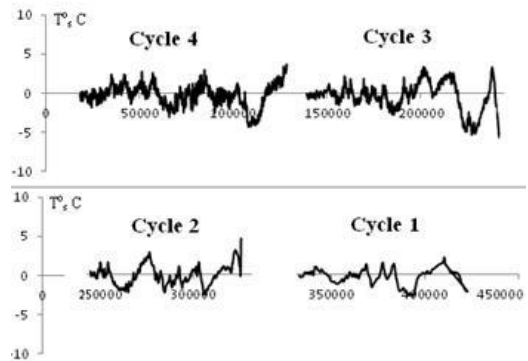

Figure 5

The high frequency temperature fluctuations $T_s$ in cycles 1-4.

To study the scaling of the Vostok ice-core high frequency data component, the author calculated the Fourier power spectrum in the cycle 4, Figure 6.

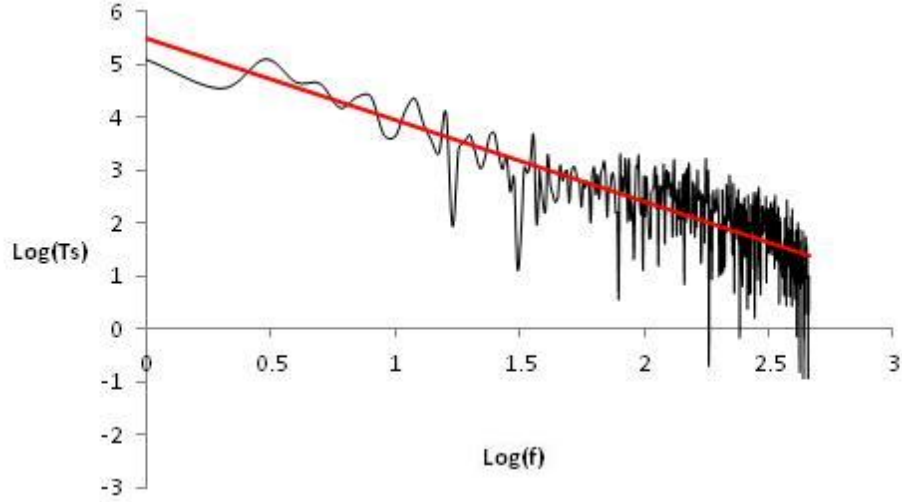

Figure 6

The power spectrum of the data.

The straight line $log(T_s) = -1.55 \cdot log(t) + 5.52$ is the indication of constant scaling coefficient -*1.55*. This coefficient is close to *β = -1.4* found in the work (Schmitt et al, 1995) for Greenland. The coefficient of determination for this approximation is $R^2=0.54$.

The power spectrum calculated above is the second order moment. To calculate the spectrum for a range of moments, we will use a new variable *α* which is related to the mass exponent *τ(q)* by

$$\alpha = \frac{d\tau(q)}{dq} \qquad (5)$$

The mass exponent function $\tau(q, \delta T)$ in (5) is

$$\tau(q, \delta T) = \frac{\ln D(q, \delta T)}{\ln(\delta T)}, \quad (6)$$

where $D(q, \delta T)$ is the partition function

$$D(q, \delta T) = \sum_{i}^{N} n_i^q, \quad (7)$$

$n_i$ is taken from a histogram of the temperature distribution with temperature discretization $\delta T$. We used two levels of discretization $\delta T = 0.25$, and 0.125 to calculate the multifractal spectrum of $T_s$. Figure 7 shows histograms for $\delta T = 0.25$ and 0.125.

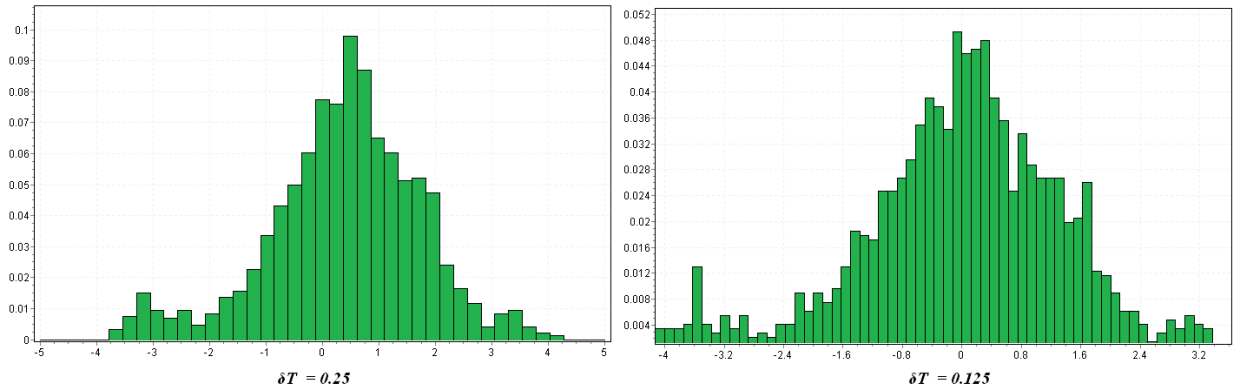

Figure 7

Histograms $n_i$ for $\delta T = 0.25$ and 0.125.

The fractal dimension $f(\alpha)$ of a subset $\alpha$ was calculated for $\delta T = 0.25$ and 0.125 using the formula (Feder, 1988):

$$f(\alpha) = q \cdot \alpha - \tau(q) \quad (8)$$

Figure 8 shows the multifractal spectrum of temperature fluctuations $T_s$ in cycle 4 (crosses) for $\delta T = 0.125$.

To model the high frequency stochastic temperature fluctuations we used a sine-circle map (9):

$$t_{n+1} = t_n + \Omega - \frac{k}{2\pi}\sin(2\pi t_n) \quad (\text{mod } 1) \quad (9)$$

This map is known for its successful modeling of non-linear dynamical systems. It has the remarkable property of producing periodic, quasi-periodic and chaotic oscillations. An extensive review of sine-circle map applications can be found in (Glazier, Libchaber, 1988). The function $f(\alpha)$ was calculated for $\Omega = \Omega_{gm} = (\sqrt{5}+1)/2$, the golden mean, $k = 1$, and plotted on Figure 8.

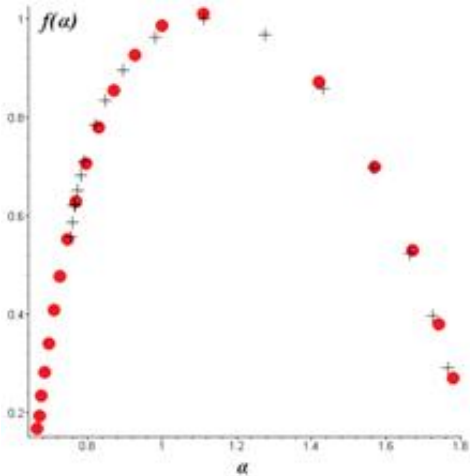

Figure 8

Multifractal spectrum of temperature variation $T_s$ in cycle 4 (crosses), and multifractal spectrum of critical circle map (circles), $\delta T = 0.125$.

## 5. Conceptual Interpretation of Models

*Cyclic temperature variations $T_c$.*

It is show in section 3 that the system of non-linear differential equations (3) has a cycle. A.Andronov in his work (1929) demonstrated the relation between the limit cycle and a special form of oscillation which is called auto-oscillation (self-oscillation, self sustained oscillation). Auto-oscillation is the self-organized response of a non-linear dynamical system to a constant, non-oscillating, flow of energy. Thus, dynamical systems, described by non-linear differential equations (3) can be considered to be a non-linear, dissipative, and self-organized system. According to Figure 4b, the peak of entropy precedes the peak of the temperature, and a decrease in entropy is followed by a rapid increase in temperature of about $10^oC$ on the Earth's surface. Physically, this can be interpreted as a phase transition, like the condensation of water vapor, and release of latent heat, as discussed it in section 2. We can see that the duration of observed temperature cycles, Figure 4a, gradually increases from ≈100ky in cycles 1 and 2, through ≈120ky in cycle 3, and to ≈130ky in cycle 4. This increase in the duration of temperature cycles can be caused by gradual changes in properties of the system itself. To model this situation, the original system of equations (3) was modified by replacing the constant coefficient *$a_1$* with coefficient *$a_1(1+αt)$, t* - time. Calculations were made for *$a_1$ = -4.673*, and for *α = ± 0.05*. The case of positive *α* corresponds to an increased rate of cooling of the system, and the case of negative *α* corresponds to a decreased rate of cooling of the system relative to that rate in a cycle. The increased rate of cooling of the system caused a gradual increase in the cycle's duration, Figure 5a, and, vice versa, for a slower rate of cooling we observe a decrease in the duration of cycles, Figure 9b.

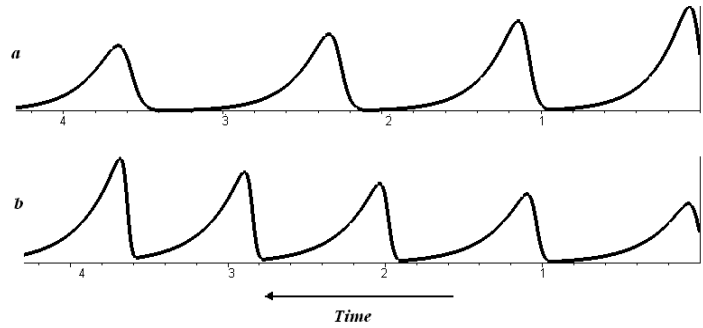

Figure 9

The influence of the changing rate of cooling on cycle duration.

*a*, increased duration of cycles caused by increased rate of cooling **-4.673(1+0.05t)**

*b,* decreased duration of cycles for slower rate of cooling **-4.673(1-0.05t)**

This result shows that the changing rate of cooling can be responsible for the duration of interglacial cycles observed in the Quaternary period.

*High frequency temperature fluctuations $T_s$.*

Figure 8 shows that the spectrum of stochastic temperature fluctuations has a multifractal structure. This spectrum is identical, within the limits of this numerical experiment, to the spectrum of the critical circle map, which the author considers a mathematical model of stochastic temperature fluctuations $T_s$. Formula (9) for $k = 1$ and irrational $\Omega$ represents quasi-periodic oscillations, the intermediate process between the cycle and chaos in the evolution of a structure from simple to complex. Theoretical and experimental studies of convection in liquids (Stavans, 1987; Glazier, Libchaber, 1988) reveal the fundamental similarity between temperature variations in convecting fluids and quasi-periodic oscillations described by the critical sine circle map (9). Thus, based on the universality of the critical circle map as a mathematical model, one can suggest that the stochastic component $T_s$ of temperature fluctuations in Antarctica is caused by convection in Earth's overheated atmosphere due to global warming. This does not contradict the hypothesis made in Section 2 for calculating the exponential trend in observed temperature. Figure 10 shows how the range of stochastic fluctuations $T_s$ changes with a change in the level of cyclic temperature $T_c$.

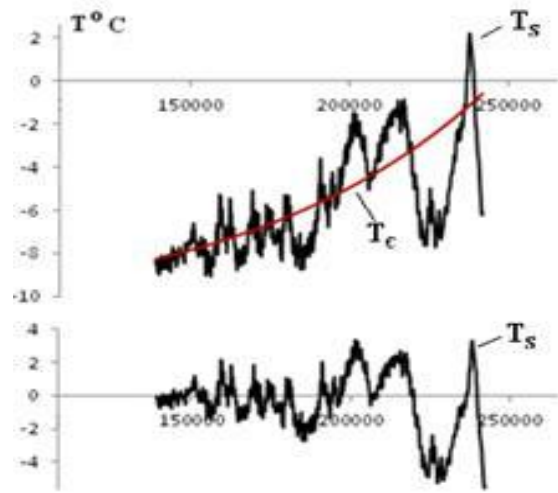

Figure 10

The cyclic $T_c$ and stochastic temperatures $T_s$ in cycle 3.

With decreasing level of $T_c$ the range of $T_s$ fluctuations decreases also. The source that restores the critical temperature threshold in the Earth's climate system is the input of solar radiation. Temperatures above this critical level trigger turbulent convection in the planetary atmosphere, cooling down the planet. We can interpret $T_c$ as the "system energy reservoir". Fluctuations, or "avalanches" of temperature $T_s$, take energy from this reservoir. With a decreasing level of energy supply, the range of temperature fluctuations decreases almost to zero. With each sharp rise of temperature in the auto-oscillating component, a series of temperature "avalanches" continues into the next cycle.

## 6. Discussion

We will now examine the results of the two processes, cyclic and stochastic, on the climate system as a whole. Looking at one temperature cycle, and starting at the interglacial period, we can propose the following conceptual model of climate evolution: as the atmosphere warms, the permafrost thaws and releases large quantities of methane. Increasing global temperature also means higher concentrations of water vapor in the atmosphere. Both of these processes intensify global warming, which gives rise to turbulent convection in the atmosphere. The result of this convection is two-fold. On the one hand, the atmosphere will begin to cool, exhibiting an exponential decay of temperature. At the same time, the amount of dust in the atmosphere increases. These two processes combine as the temperature cools enough for water vapor to condense on the dust particles. This release of latent heat will result in an increase of temperature on the Earth's surface, while, at the same time, the atmosphere will clear, and a new cycle will begin.

## 7. Conclusion

It is shown that the temperature variation data in Antarctica can be represented as the sum of two parts, conventionally called the "cyclic" and "high frequency, stochastic" components. These two components are evidence of two different, but tightly interconnected, global climate processes. The first one is the sequence of temperature cycles, with periods gradually increasing from 100,000 years in the first cycle to approximately 130,000 years in the last glacial cycle. The second process is the high frequency fluctuation of temperature in each cycle. The self-organization in the auto-oscillation process is the non-linear reaction of the Earth's climate system, as a whole, to the input of solar radiation. The self-organization in the high frequency part is the self-organized nonlinear critical process taking energy from, and fluctuating around the auto-oscillating part of the temperature variations. These models characterize the Earth's climate as an open, complex, self-organized, dynamical system with nonlinear reaction to the input of solar radiation. The solar activity and variations in Earth's orbital parameters are external factors that can be taken into account as forcing functions in the dynamical model of climate. To model the actual data as close as possible, we consider solving the system of equations (3) with time-dependent coefficients and a forcing function which represents insolation. The second direction of our continued research is the study of the high-frequency component $T_s$ of temperature decomposition as a non-stationary multifractal time series.

**Acknowledgments**

The author would like to thank the anonymous reviewers for their valuable comments and suggestions.